 \documentclass[final,5p,times,authoryear]{article}

\usepackage{amssymb}
\usepackage{lipsum}
\usepackage{multirow}
\usepackage{graphicx}
\usepackage{authblk}

\begin{document}


\title{Towards polarization-enhanced PET:\\
Study of random background in polarization-correlated Compton events}

\author[1,2]{Ana Marija Kožuljević}
\author[1]{Tomislav Bokulić}
\author[3]{Darko Grošev}
\author[4]{Siddharth Parashari}
\author[2]{Luka Pavelić}
\author[5,6,7]{Marinko Rade}
\author[3]{Marijan Žuvić}
\author[1]{Mihael Makek}

\affil[1]{{University of Zagreb, Faculty of Science, Zagreb, Croatia}}
\affil[2]{{Institute for Medical Research and Occupational Health, Zagreb, Croatia}}
\affil[3]{{University Hospital Centre Zagreb, Zagreb, Croatia}}
\affil[4]{{Instituto de Física Corpuscular (IFIC), CSIC-UV, Valencia, Spain}}
\affil[5]{{Kuopio University Hospital, University of Eastern Finland, Kuopio, Finland}}
\affil[6]{{Juraj Dobrila University, Pula, Croatia}}
\affil[7]{{Medi-lab d.o.o., Zagreb, Croatia}}


\maketitle
\begin{abstract}
\textbf{Background:} Positron Emission Tomography (PET) is a medical imaging modality that utilizes positron-emitting isotopes, such as Ga-68 and F-18, for many diagnostic purposes. The positron annihilates with an electron from the surrounding tissue, creating two photons of 511 keV energy and opposite momenta, also entangled in their orthogonal polarizations. If each photon undergoes a Compton scattering process, the difference of their azimuthal scattering angles reflects the initial orthogonality of the polarizations, peaking at $\pm$90$^{\circ}$. This correlation, not yet utilized in conventional PET scanners, offers an additional, energy-independent method for background discrimination.
\textbf{Methods:} It can be measured using Compton polarimeters, and particularly suitable are single-layer polarimeters, compatible with conventional PET architecture. We assembled demonstrator with two such modules comprising $3 \times 3 \times 20$ mm$^3$ GAGG:Ce scintillating pixels in $16\times16$ matrix, read by silicon photomultipliers, mounted on a rotating gantry with 430 mm diameter.
\textbf{Results:} This paper reports on the study of random background in measurements with Ga-68 source with activities $200-380$ MBq. We compare a sample of polarization-correlated Compton events, having a two-hit signature, to a sample of conventional single-pixel hits. We find that the signal-to-random background ratio obtained in the polarization-correlated events is larger than the one for the single-pixel hits, for all selected event samples. We also demonstrate a correlation between the signal-to-random background ratio and the polarimetric modulation factor $\mu$.
\textbf{Conclusion:} The random background suppression in the measurements of the polarization-correlated annihilation quanta is higher than in the standard PET modality, which could be a valuable resource for PET imaging. Since there is a correlation between the signal-to-random background and the polarimetric modulation factor, it could serve as an estimator of the random background.
  
\end{abstract}


Positron Emission Tomography, quantum entanglement, photon polarization, random background, polarimetric modulation


\section{Introduction}\label{introduction}

Recently, a new path for enhancing Positron Emission Tomography (PET) has been proposed, to utilize the correlation of polarization of annihilation photons \cite{kuncic2011, Toghyani2016, Watts2021, Moskal2024a}. Quantum mechanics, through parity and spin conservation, necessitates the entanglement of the two annihilation photons, reflected in orthogonality of their polarizations \cite{pryce} - a property not yet utilized in PET, which conventionally measures only their 511 keV energies and opposite momenta. To create a line of response (LOR) through the annihilation site, the two photons originating from it must be detected in coincidence. Since the timing window is finite, it opens an opportunity for the detection of two uncorrelated photons, creating a random coincidence event. Such a LOR does not correspond to the annihilation site and, consequently, degrades the quality of the image. The information about the entanglement of the annihilation photons, exploited by measurement of their polarization correlations, is therefore an independent tool to reduce the random coincidence background, since it lacks this type of correlation \cite{kuncic2011, Toghyani2016, Watts2021, Mcnamara2014, Kim2023}. 

The polarization correlations of annihilation photons can be measured through the Compton scattering process \cite{pryce, Makek2020}(Figure \ref{fig_schema}). If both annihilation photons detected in coincidence undergo Compton scattering in their respective detectors, through measurements of the Compton scattering angle $\theta$ and azimuthal scattering angle $\phi$, it is possible to deduce the azimuthal scattering angle difference $\Delta\phi$. The $\Delta\phi$ values reflect the initial orthogonality of the polarizations (Figure \ref{fig_mu}), with the largest cross-section for the $\pm$90\textdegree difference for fixed values of $\theta_{1,2}$ \cite{pryce}:

\begin{equation}
    \mathrm{ \frac{d^2\sigma}{d\Omega_1\!d\Omega_2}\!\!=\!\!\frac{r_0^4}{16}\left[F(\theta_1\!)F(\theta_2\!)-\!G(\theta_1\!)G(\theta_2\!)cos(2\Delta\phi)\right] },
\end{equation}
with $F\left(\theta_{i}\right) = \frac{\left[2+\left(1-\cos \theta_{i}\right)^{3}\right]}{\left(2-\cos \theta_{i}\right)^{3}}$ and $G\left(\theta_{i}\right) = \frac{\sin ^{2} \theta_{i}}{\left(2-\cos \theta_{i}\right)^{2}}$, where $\mathrm{r_0}$ is the classical electron radius, $\mathrm{d\Omega_{1,2}}$ are the solid angles, $\mathrm{\theta_{1,2}}$ are the Compton scattering angles, and $\mathrm{\Delta\phi\!=\!(\phi_1-\phi_2})$ is the difference of the azimuthal scattering angles of photon 1 and 2, respectively. The polarimetric modulation factor $\mu$:
 \begin{equation}
 \mu = G(\theta_1)G(\theta_2)
\label{eqn_mu}
\end{equation}
reflects the sensitivity of the detector setup to the initial orthogonality of the polarizations. This factor will have the highest value of $\sim$42\% for $\theta_{1,2}\approx$82\textdegree\; for an ideal detector \cite{Toghyani2016} in the absence of background. The experimentally determined values will always be lower than the maximum due to the finite detector acceptance. In addition, the measured value of $\mu$ could depend on the environment in which the positron is emitted \cite{Kumar_2025}, and as shown in this work, on the random background rate. There is also an ongoing experimental \cite{ivashkin_testing, PARASHARI_PLB, Tkachev2025, PhysRevLett.133.132502}, as well as theoretical discussions about the precise description of the conducted measurements and the nature of the annihilation process \cite{CARADONNA2024169779, Cardonna_2025_physrevA, cardonna_shimazoe, beatrix_scirep, beatrix_2025}.

Various concepts of utilizing polarization correlations to improve event classification in PET have been investigated experimentally \cite{Watts2021, Parashari2022_nima, Kim2023, Moskal_2025_sa} under specific laboratory conditions. This work continues the exploration of the polarization-sensitive PET  concept, building upon the initial results obtained with the demonstrator setup \cite{Makek2022, Kozuljevic2024}. It presents the first detailed comparative study of the random background in single-pixel hit events, and correlated Compton events.

\section{Methods}\label{Methods}

\subsection{The detector modules}
\begin{figure}
	\centering 
	\includegraphics[width=0.8\textwidth]{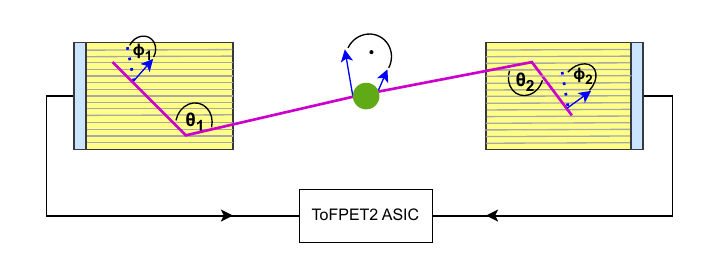}	
	\caption{Simplified schematics of the experimental setup. The figure represents two segmented scintillating GAGG detector modules (in yellow) with SiPMs (in blue) in a coincidence measurement setup, read-out by the ToFPET2 ASIC system. The source (in green) emits two polarization-correlated annihilation photons (in purple) that undergo Compton scattering in their respective detector modules with Compton scattering angles $\theta_{1,2}$ and azimuthal scattering angles $\phi_{1,2}$. } 
	\label{fig_schema}%
\end{figure}

Two detector modules were assembled from four 8$\times$8 matrices of GAGG:Ce crystals, with each crystal sized at 3 mm $\times$ 3 mm $\times$ 20 mm, totaling two 16$\times$16 detector matrices with a 3.2 mm matrix pitch. The SiPMs (Hamamatsu Photonics, model S13361-0808A) were glued to the crystals in a one-to-one coupling with optical cement (EJ-500, Eljen Technology, USA). The relative energy resolutions at 511 keV of the two detectors were found to be (8.1$\pm$1.1)\% and (9.3$\pm$2.2)\% \cite{Parashari2022_nima}. The detector modules were mounted on a rotating gantry and arranged so that their faces opposed each other at a distance of 430 mm, with a container of the Ga-68 solution placed between the modules at the center of the gantry's ring (Figure \ref{fig_schema}).

\subsection{Data acquisition}
Measurements were conducted at the Department of Nuclear Medicine and Radiation Protection, University Hospital Centre Zagreb, with a Ga-68 isotope source with an initial activity of $\sim$378 MBq. The read-out and digitization of the data acquired in this setup were performed with the TOFPET2 ASIC system (PETsys Electronics — Medical PET Detectors, S.A., Oeiras, Portugal)\cite{Francesco2016}. It was set to trigger on coincidences in which at least one pixel fired per module. Each such hit had a timestamp that allowed grouping the hits within 100 ns into events. In case of the events in which two pixels fired in a module, only the pixel of the first interaction was considered. The pixel that is considered to have fired first is the one with lower deposited energy \cite{Kozuljevic2021}. The difference in timestamps for events in which two pixels fired in opposite detectors can be seen in Figure \ref{fig_spectra}, which represents plots of coincidence time spectra at the highest activity. The duration of the experiment was approximately one hour, and the duration of acquisition at each of the 12 positions around the source (15\textdegree\; apart) is given in Table \ref{tab_acquisition}.

\begin{table}[]
\centering
\begin{tabular}{|c|c|c|c|c}
\hline
&&& \\
Position & \multirow{2}{2cm}{\centering Acquisition time [s]} & Total time [s] & Activity [MBq] \\
& & & \\
&&& \\
\hline
1 & 240 & 260 & 377.7 \\
2 & 251 & 272 & 361.4 \\
3 & 262 & 283 & 345.1 \\
4 & 275 & 297 & 328.9  \\
5 & 289 & 311 & 312.7  \\
6 & 304 & 326 & 296.6  \\
7 & 322 & 345 & 280.6  \\
8 & 341 & 364 & 264.6  \\
9 & 362 & 387 & 248.8  \\
10 & 387 & 412 & 232.9  \\
11 & 415 & 441 & 217.2 \\
12 & 447 & 475 & 201.5 \\
\hline
\end{tabular}
    \caption{The experiment was conducted by rotating the detector modules around the Ga-68 source, and the data were acquired in 12 positions. The length of acquisition at each position is given in the Table, as well as the total time (acquisition length with added time required for rotation to the next position) and activity of the Ga-68 source at the start of each acquisition. }
    \label{tab_acquisition}
\end{table}

\subsection{Data selection and analysis}
\begin{figure}
	\centering 
	\includegraphics[width=0.8\textwidth]{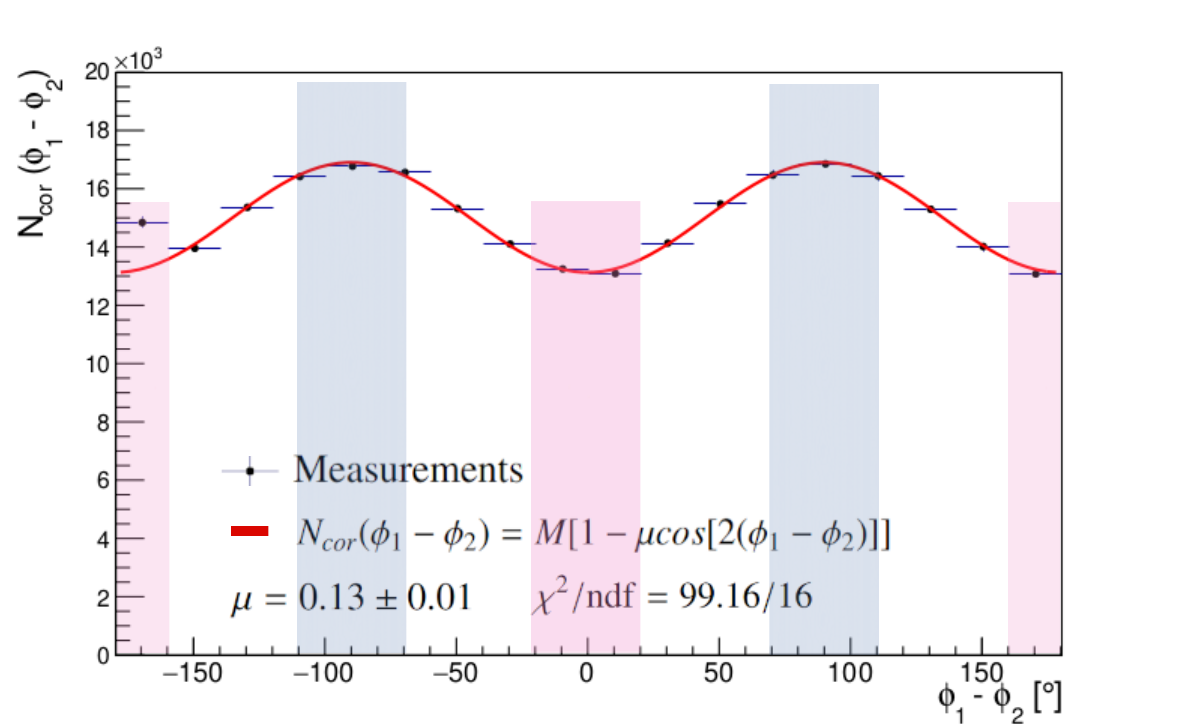}	
	\caption{Acceptance corrected $\phi_1$-$\phi_2$ distribution at the sixth measurement position. The Compton scattering angle range was selected as $72^{\circ}<\theta<90^{\circ}$. The highlighted regions denote the azimuthal difference range of events selected for further analysis. The regions shaded in blue correspond to the $70^{\circ}<\Delta\phi<110^{\circ}$ , and denote the highest probability of polarization correlations, while the regions shaded in pink, with $-20^{\circ}<\Delta\phi<20^{\circ}$ , denote the lowest probability of two photons having correlated polarizations.} 
	\label{fig_mu}%
\end{figure}

From the collected events, two types of events were selected: the single-pixel events, where each detector module fired with only one pixel in which the full energy (from 400 keV to 600 keV) of the annihilation photon was deposited (i.e., photoelectric events), and events with two pixels fired per module with the sum of the deposited annihilation photon energies in the full-energy range, with the minimum deposited energy of 120 keV per pixel. From the energies of the two pixels, Compton scattering angles $\theta_{1,2}$ can be determined:
\begin{equation}
         \theta \!=\! \mathrm{acos\left(\frac{m_ec^2}{E_{px_1}\!\!+\!E_{px_2}} \!-\! \frac{m_ec^2}{E_{px_2}}\!+1\!\right)};
     \label{eqn:angles}
\end{equation}
and from the topology of the detector modules, the azimuthal scattering angle $\phi$:
\begin{equation}
     \phi \!=\! \mathrm{atan}\left(\frac{\Delta y}{\Delta x}\right)
\end{equation}
The measured distribution of the difference of the azimuthal scattering angles $\Delta\phi = \phi_{1}-\phi_{2}$ is influenced by the distance of the two pixels that fired $d=\sqrt{(\Delta x)^{2}+(\Delta y)^{2}}$, since the detector has non-uniform acceptance for different event topologies, with events in which $d$ is very large being almost negligible. This acceptance effect is corrected by normalizing the obtained distribution according to: 
\begin{equation}
    N_{cor}(\phi_1 - \phi_2)=\frac{N(\phi_1 - \phi_2)}{A_{uncor}(\phi_1 - \phi_2)}.
\end{equation}
where $A_{uncor}$ is obtained from uncorrelated events \cite{Makek2020}. 
The corrected distribution $N_{cor}$ allows the quantification of the polarimetric modulation factor $\mu$ by fitting the Pryce-Ward-shaped relation:
\begin{equation}
    N_{cor}(\phi_1 - \phi_2)=M [1 - \mu \mathrm{cos}[2(\phi_1 - \phi_2)]], 
\end{equation}
as can be seen in Figure \ref{fig_mu}. The more in-depth description of the analysis process can be found in \cite{Makek2020, Kozuljevic2024, Parashari2022_jinst}.

\begin{figure}
	\centering 
	\includegraphics[width=0.8\textwidth]{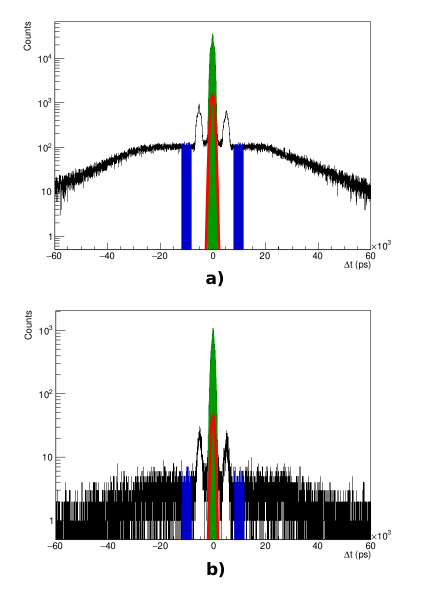}	
	\caption{Coincidence time spectra for a) single-pixel (photoelectric) events and b) polarization-correlated Compton events. The selected signal (in green) is estimated as $\pm$3$\sigma$ from the mean calculated by the Gaussian fit (red curve) to the highest peak in each spectrum. The random background (in blue) is estimated as 6$\sigma$ area, removed from the calculated mean by $\pm$15$\sigma$.} 
	\label{fig_spectra}%
\end{figure}

The coincidence time spectra for measurements at the highest activity (first position) are presented in Figure \ref{fig_spectra}. The central peak was fitted by a Gaussian to determine its mean and standard deviation, and the data selected within $\pm$3$\sigma$ from the peak centroid (i.e. $\sim\pm$2.5 ns from the peak) was considered as a signal. The random background data was chosen as a 6$\sigma$ wide area at $\pm$15$\sigma$ ($\sim\pm$ 8.5 ns) distance from the calculated peak. The small off center peaks are a know artifact of the TOFPET2 ASIC \cite{tofpet2_evaluation}. The regions of interest are chosen to avoid them. After normalization of the random background data, the signal-to-random background (SBR) ratio is calculated by dividing the number of selected events in the signal and random background areas.

The data analysis was conducted in ROOT \cite{ref-root}, with the optimization help from the GNU \textit{parallel} command \cite{ref-parallel}. The obtained data was plotted in Python.

\section{Results}\label{Results}

\begin{figure}
	\centering 
	\includegraphics[width=0.8\textwidth]{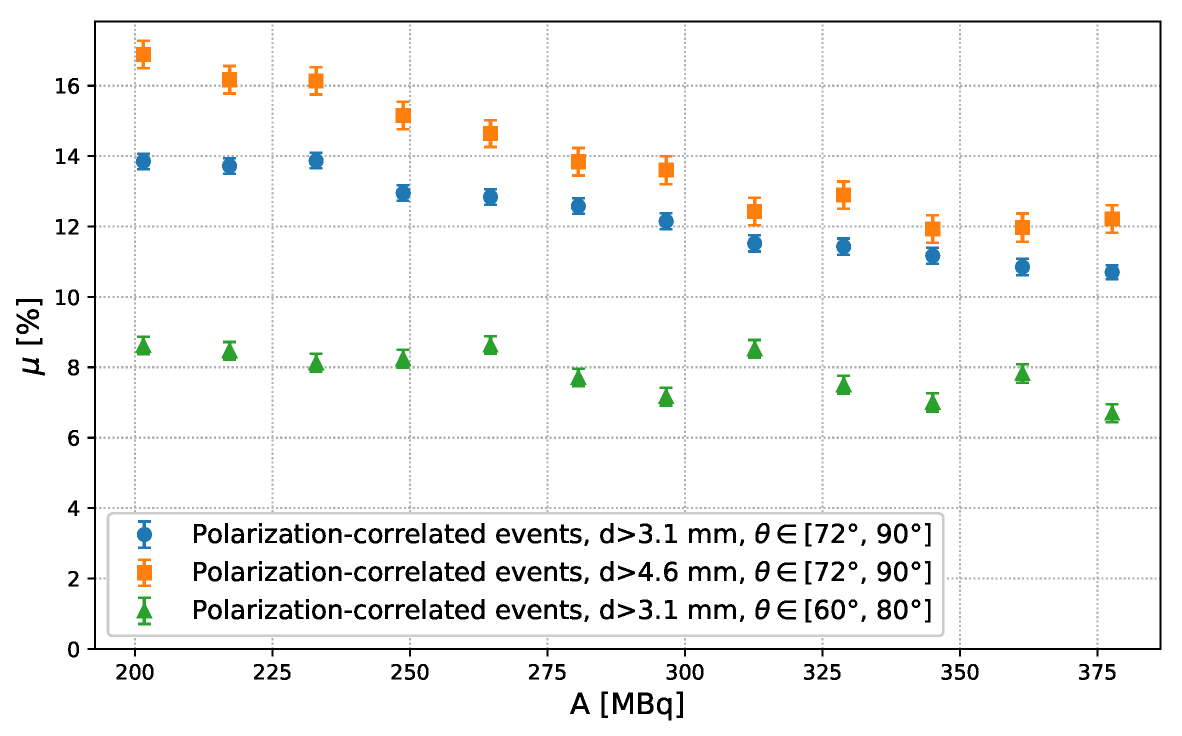}	
	\caption{The measured values of the polarimetric modulation factor $\mu$ exhibit higher values at lower source activity A levels for all selected event topologies.} 
	\label{fig_muvsa}%
\end{figure}

\begin{figure}
	\centering 
	\includegraphics[width=0.8\textwidth]{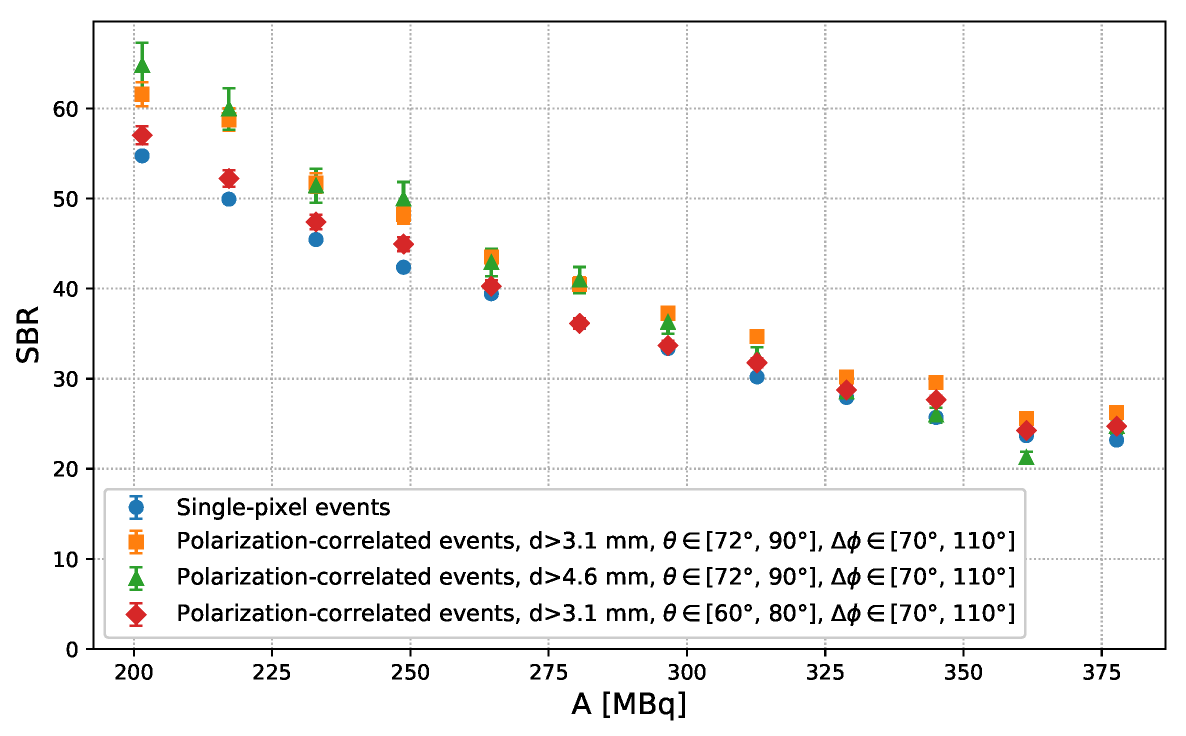}	
	\caption{The variation of SBR of different polarization-correlated events' topologies compared to the single-pixel events with the source activity A.
    } 
	\label{fig_multisbrvsa}%
\end{figure}

We first investigated how the polarimetric modulation factor $\mu$ behaves during the measurements with this experimental setup at high (approx. 200 MBq - 378 MBq) activity levels. The results are presented in Figure \ref{fig_muvsa}. For angular selection of $72^{\circ}<\theta<90^{\circ}$, we considered two samples based on Compton event topology, one with any pair of fired pixels, the other containing pairs with the nearest neighbors excluded (d \textgreater\;4.6 mm). A third sample with $60^{\circ}<\theta<80^{\circ}$, and all pixel pairs is added for comparison. The first sample exhibits $\sim$28\% increase in $\mu$ as the source activity declines from $\sim$387 MBq to $\sim$200 MBq. The second,  shows higher $\mu$ values than the former, which is expected \cite{Makek2017, Parashari2022_jinst} owing to better angular precision. The selection also has a gain of $\sim$40\% at the end of the measurement. The third sample demonstrates the lower polarimetric sensitivity, as expected from eq. \ref{eqn_mu}, with a maximum $\mu\sim$8.5\%. 
In addition, we explored how the SBR values of these samples vary with the source activity. The results can be seen in Figure \ref{fig_multisbrvsa}. We observe a clear gain in SBR at the lower activities, however the difference beween the samples is not pronounced.  

\begin{figure}
	\centering 
	\includegraphics[width=0.8\textwidth]{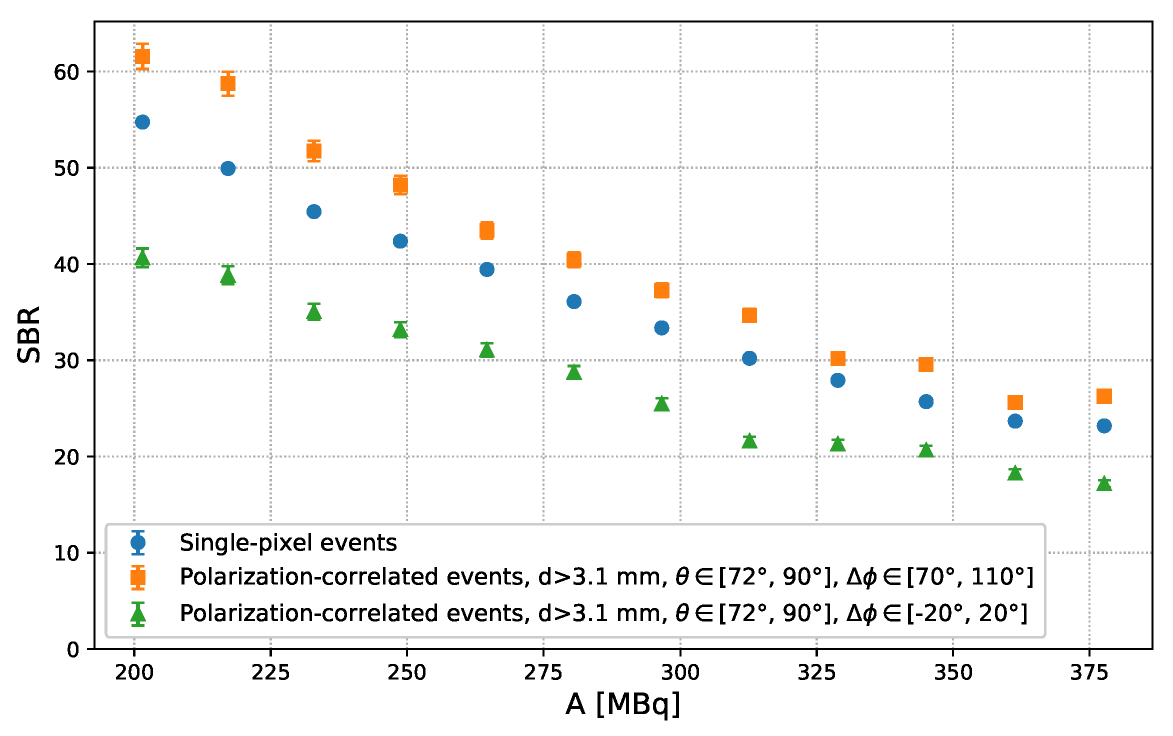}	
	\caption{The SBR variation with the source activity, A, for different event samples.}
	\label{fig_sbrvsa}%
\end{figure}

We also examined how the signal-to-random background ratios, calculated from the coincidence time spectra (Figure \ref{fig_spectra}), varied with the source activity for single-pixel events and different selections of polarization-correlated events. In Figure \ref{fig_sbrvsa}, the events with polarization-correlated quanta at $72^{\circ}<\theta<90^{\circ}$ and $70^{\circ}<\Delta\phi<110^{\circ}$ (i.e. the blue-shaded area in Figure \ref{fig_mu}) exhibit higher random background suppression compared to the single-pixel events and the polarization-correlated events with the same $\theta_{1,2}$ selection; however, the selection of the $\Delta\phi$ that encompasses the valleys of the distribution (areas shaded in pink in Figure \ref{fig_mu}) shows the lowest SBR of all examined event types. 

\begin{figure}
	\centering 
	\includegraphics[width=0.8\textwidth]{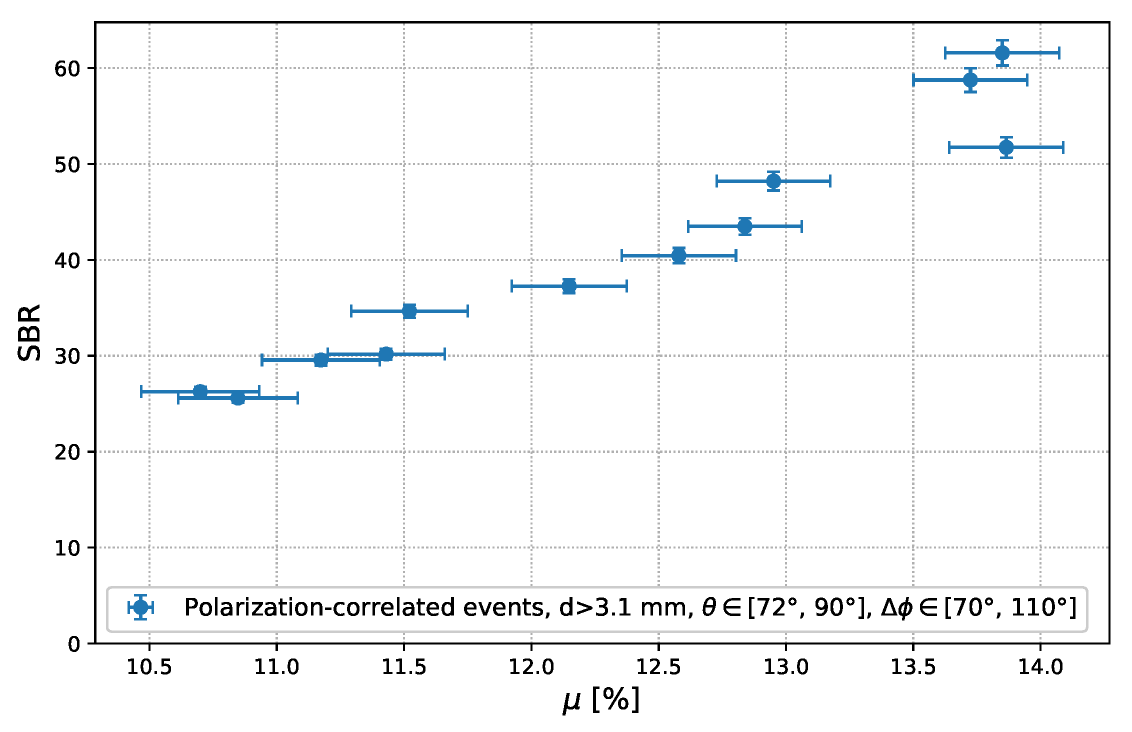}	
	\caption{The variation of the SBR with the modulation factor $\mu$, for events with $72^{\circ}<\theta<90^{\circ}$.} 
	\label{fig_sbrvsmu}%
\end{figure}

The SBR dependence of the polarization-correlated events on the polarimetric modulation factor $\mu$ has been extracted and shown in Figure \ref{fig_sbrvsmu}. It demonstrates an increase of the SBR by a factor of $\sim2$ with $\mu$ variation in $10.5\% < \mu < 14\%$ range. This signals that the polarimetric modulation is a clear indication of the random background contents of the sample. 

\begin{figure}
	\centering 
	\includegraphics[width=0.8\textwidth]{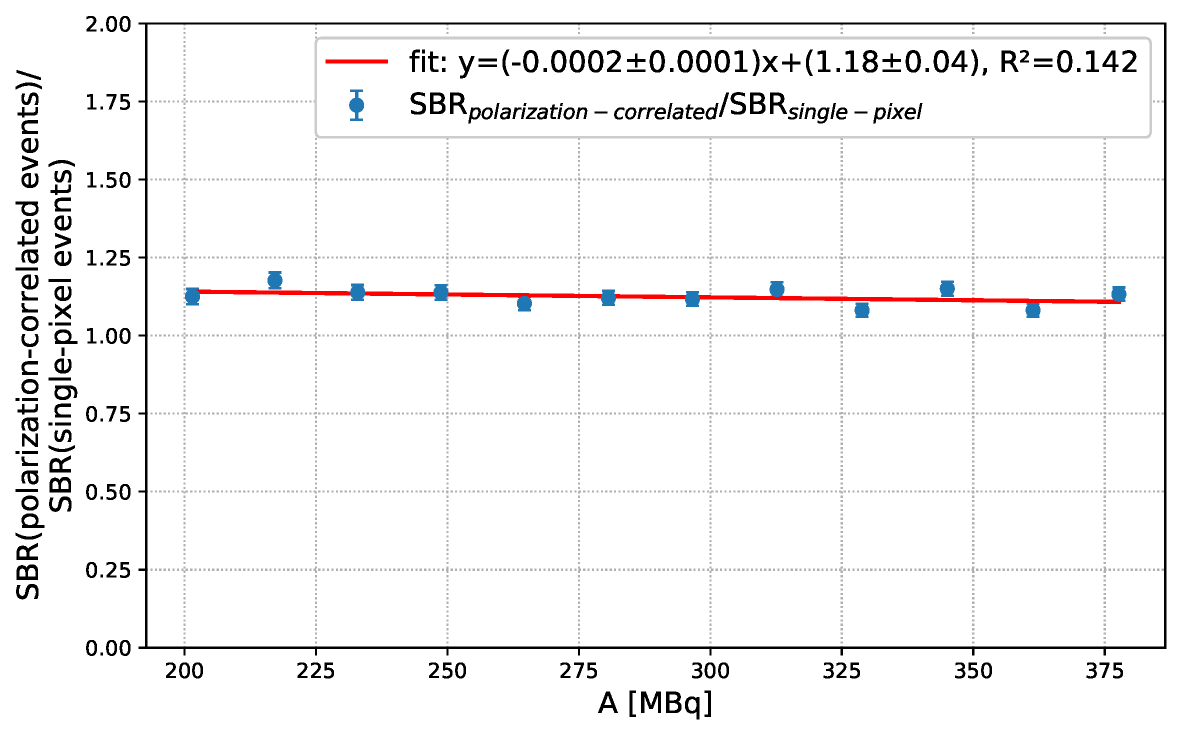}	
	\caption{The ratios of the SBR values obtained with the polarization-correlated events and single-pixel events and their linear dependance on the source activity A. } 
	\label{fig_ratio}%
\end{figure}

Finally, we extracted the ratio of the SBR values for selected polarization-correlated events and single-pixel events \ref{fig_ratio}. We observe a falling trend with the ratio being reduced at higher activities, but remains larger than 1, favoring polarization-correlated events in this respect.

\section{Discussion and conclusion}
We performed an experimental study using a polarization-sensitive PET demonstration setup. Data from a Ga-68 source were acquired, with activity in 200-380 MBq range. Polarization-correlated events, with specific two pixel topologies were selected, along with a sample of single-pixel events.

For the polarization correlated Compton events, we demonstrate a higher SBR for any selected event topology, compared to single-pixel events. For the largest selected sample of those events ($72^{\circ}<\theta<90^{\circ}$ and azimuthal angle difference $70^{\circ}<\Delta\phi<110^{\circ}$), this ratio remains above 1.1, suggesting that the polarization-correlated events may be a valuable resource for PET imaging. We also observe a raising trend of this ratio with lower activities, pointing at further enhancement. 

This study also reveals for the first time the dependence of the signal-to-random background ratio on the polarimetric modulation factor. This clear correlation indicates that $\mu$ can be an additional estimator of the random background, complementary to single rate-based or delayed window-based random estimates in PET imaging. The dependence of the $\mu$ on the random background has also implications on proposed polarization-sensitive positronium imaging \cite{Moskal_2025_sa}, where it needs to be disentangled from the positronium formation effects.

\section*{Acknowledgements}
This work was supported by the “Research Cooperability” Program of the Croatian Science Foundation, funded by the European Union from the European Social Fund under the Operational Programme Efficient Human Resources 2014–2020, grant number PZS-2019-02-5829. This work was also supported by the Croatian Science Foundation under the project number IP-2022-10-3878.

\bibliographystyle{unsrt} 
\bibliography{bibliography}






\end{document}